\documentclass[prl,aps,twocolumn,showpacs]{revtex4}
\usepackage{epsfig}
\usepackage{graphicx}
\usepackage{textcomp}

\begin{document}
\title{WKB theory of epidemic fade-out in stochastic populations}

\author{Baruch Meerson$^{1}$ and Pavel V. Sasorov$^{2}$}

\affiliation{$^{1}$Racah Institute of Physics, Hebrew University
of Jerusalem, Jerusalem 91904, Israel}

\affiliation{$^{2}$Institute of Theoretical and Experimental
Physics, Moscow 117218, Russia}


\begin{abstract}
Stochastic effects may cause `fade-out'
of an infectious disease in a population
immediately after an epidemic outbreak. 
We evaluate the epidemic fade-out probability by a WKB method and find that
the most probable path to extinction of the disease comes from
an instanton-like orbit in the phase space of an underlying Hamiltonian flow.

\end{abstract}

\maketitle

An infectious disease can disappear from a population immediately after a major infection outbreak \cite{Bartlett,vH2}.
This phenomenon, called ``epidemic fade-out", occurs if the epidemic dynamics
is oscillatory, and the number of infected individuals at the end of the first outbreak of the disease
is relatively  low so that
fluctuations in the disease transmission can ``switch off" the disease.
Epidemic fade-out
has been addressed by epidemiologists via stochastic simulations. One
exception is Ref. \cite{vH2} which  arrived at important analytical results which we
briefly review and use below.

Epidemic fade-out is an example of a large fluctuation
in a multivariate stochastic system far from equilibrium. There is no
general theory of fluctuations in such systems, and finding the probability
of a large fluctuation is hard. Here we develop a theoretical framework for analysis of
epidemic fade-out of the example of
stochastic SI model: a Markov process involving
Susceptible and Infected sub-populations \cite{Bartlett,vH2,SIR}.  We
formulate the problem in a master equation setting. In contrast
to \textit{endemic} fade-out, which can be studied assuming
quasi-stationarity \cite{KM}, the epidemic fade-out occurs
on a \textit{fast} time scale (determined by the deterministic rate equations of the SIR model), so no ready-to-use methods of solution are available.
To overcome this difficulty we derive a \textit{stationary}
equation for the mean residence time of the population in a certain state.
Then we develop a WKB theory using the population size as a large parameter. In the WKB framework the problem reduces to
that of finding a special zero-energy phase orbit of the underlying  Hamiltonian: the orbit
which provides a global minimum to the
action functional for boundary conditions, corresponding to epidemic fade-out.
This orbit, which encodes the most probable path of the population towards the disease extinction, turns out to be instanton-like \cite{instanton}.
We observe that the epidemic fade-out instanton exists only
in the regime when the epidemic dynamics, as described by the deterministic rate equations, is oscillatory. Of special interest is the regime where the
number of infected exhibits \textit{large} oscillations
prior to reaching the endemic state. By using a matched asymptotic expansion,
we analytically calculate the action along the instanton which determines
the epidemic fade-out probability.

The probability $P_{n,m}(t)$ to observe, at time $t$, $n$ susceptible and $m$ infected individuals is
governed by the master equation with transition rates from Table 1. Solving the master equation analytically is hard. The often used
van Kampen system-size expansion (vKSSE) \cite{Gardiner} approximates the master equation
by a Fokker-Planck equation. In the context of epidemic fade-out in the SI model the vKSSE was employed
in Ref. \cite{vH2}.  The vKSSE is very useful for ``typical" fluctuations \cite{Gardiner}, but it often fails for large fluctuations \cite{gaveau}, such as those causing extinction.

\begin{table}[ht]
\begin{ruledtabular}
\begin{tabular}{|c|c|c|}
 Event & Type of transition &  Rate\\
  \hline
  Infection & $S\to S-1, \, I\to I+1$ &  $(\beta/N) SI$\\
  Renewal of susceptible & $S\to S+1$ & $\mu N$ \\
  Removal of infected& $I\to I-1$ & $\Gamma I$\\
  Removal of susceptible & $S\to S-1$ & $\mu S$ \\
\end{tabular}
\end{ruledtabular}
\caption{Stochastic SI model}\label{table}
\end{table}

\begin{figure}[ht]
\includegraphics[width=2.3 in,clip=]{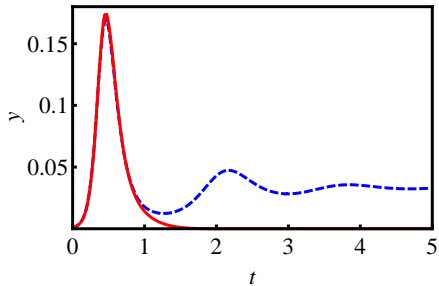}
\caption{(color online). An epidemic outbreak in the SI model. Shown is the rescaled number of
infected $y=I/N$ versus rescaled time $\mu t$.
Dashed line: prediction from the rate equations~(\ref{foi_970}). Solid line:  the
epidemic fade-out instanton. The rescaled parameters
$K=\beta/\mu=30$ and $\delta=1-\Gamma/\beta=0.5$.}
\label{y(t)}
\end{figure}

Before dealing with the master equation, consider the deterministic \textit{rate equations} for the SI model:
\begin{equation}\label{SIdot}
    \dot{S} = \mu N-(\beta/N) S\,I-\mu S\,,\;\;\;\;\;\dot{I} = (\beta/N) S\,I -\Gamma I   \,.
\end{equation}
For $\beta > \Gamma$ 
there is an attracting fixed point $\bar{S} =(\Gamma/\beta) N$, $\bar{I}=\mu (1/\Gamma-1/\beta)N$
which describes an endemic infection level, and an unstable (saddle) point $\bar{S}=N, \,\bar{I}=0$
which describes an infection-free population. At $\mu > 4 \,(\beta-\Gamma) (\Gamma/\beta)^2$
the attracting fixed point is a stable node. We are mostly interested in the
opposite inequality, when the attracting fixed point is a stable focus, and the
epidemic dynamics is oscillatory. Let
a few infected be introduced into a susceptible population. For small
$\mu$
the minimum number of infected at the end of the first outbreak of the disease is
small,  see the dashed line in Fig. \ref{y(t)}.
As a result,
stochasticity, missed by
the rate equations, can ``switch off" the disease
before the endemic level is reached. To describe this process we use the master equation
\begin{eqnarray}
\label{master}
\dot{P}_{n,m}&=& \sum\limits_{n^{\prime}, \, m^{\prime}} M_{n, \, m;\, n^{\prime}, \, m^{\prime}}\, P_{n^{\prime}, \, m^{\prime}}(t)\nonumber \\
&=&\!\mu \left[N(P_{n-1,m}\!-P_{n,m})
+(n\!+\!1)P_{n+1,m}\!-n P_{n,m}\right]\nonumber\\
&+& \Gamma\left[(m+1)P_{n,m+1}-mP_{n,m}\right]\nonumber\\
&+&\! (\beta/N)\! \left[(n+1) (m-1)P_{n+1,m-1}-nm P_{n,m}\right] .
\end{eqnarray}
A natural initial condition is a product of
Kronecker deltas: $P_{n,m}(t=0) = \delta_{n,N} \delta_{m,m_0}$.
One boundary condition reflects the fact that $m=0$ is, for any $n$,
an absorbing state. Being interested in epidemic fade-out, we
exclude from consideration all stochastic trajectories that
do not reach the extinction boundary $m=0$ immediately after the first outbreak
and leave the region of small $m$. This is achieved by introducing an artificial
absorbing boundary \cite{Gardiner} that will be specified shortly.

The disease can only disappear from the population via
transition from a state $(n,1)$ 
to the state $(n,0)$. Consider the mean residence time $T_{n,m}=\int\limits_0^\infty P_{n,m}(t)\, dt$ of the system in the state $(n,m)$, where $m>0$.  The
accumulated extinction probability ${\cal P}_n$ from the state $(n,1)$ is
${\cal P}_n =\Gamma \, T_{n,1}$, and the total extinction probability
is ${\cal P}=\sum_n{\cal P}_n$.
Integrating Eq.~(\ref{master})
over $t$ from $0$ to $\infty$ and using the
equality $P_{n,m}(t=\infty)=0$ and the initial
condition, we obtain a \textit{stationary} equation for $T_{n,m>0}$:
\begin{equation}
\sum\limits_{n^{\prime}, m^{\prime}>0} M_{n,  m;\, n^{\prime},  m^{\prime}}\, T_{n^{\prime}, m^{\prime}}
+\delta_{n, n_0}\, \delta_{m, m_0}=0\,.
\label{foi_760}
\end{equation}
We assume throughout this work that $N\gg 1$. Here the stochasticity is weak (but very
important), and Eq.~(\ref{foi_760})
can be approximately solved by the WKB ansatz $T_{n, m}=a(x,y)\, e^{-N S(x,y)}$ \cite{Kubo},
where $a$ and $S$ are smooth functions of the \textit{continuous} variables $x=n/N-1$ and $y=m/N$.

In the leading WKB order one 
arrives at a stationary Hamilton-Jacobi equation $H(x,y,\partial_x S,\partial_y S)=0$, where
\begin{eqnarray}
  &&\!\!\!\!\!\!H(x,y,p_x,p_y) = e^{p_x}-1+(1+x)\,\left(e^{-p_x}-1\right) \nonumber \\
  &&\!\!\!\!\!\!\!\!+K(1-\delta)y\left(e^{-p_y}-1\right)+K y(1+x)\left(e^{p_y-p_x}-1\right),
  \label{foi_40}
\end{eqnarray}
and we have introduced  rescaled parameters $\delta=1-\Gamma/\beta$ and $K=\beta/\mu$
and rescaled time by the rate constant $\mu$ \cite{canonical}. The four-dimensional (4d)
phase space, defined by the Hamiltonian (\ref{foi_40}), yields an instructive visualization
of the  most probable
path of the disease toward
fade-out. As $H$  is independent of time,  $H(x,y,p_x,p_y)=E=const$.
Furthermore, in view of stationarity of the Hamilton-Jacobi
equation, we only need to deal with zero-energy orbits, $E=0$.
The simplest among them are
fluctuationless orbits lying in the plane $p_x=p_y=0$. These are described by the
equations
\begin{equation}\label{foi_970}
    \dot{x} =-x - K\, y(1+x) \,,\;\;\;\;\dot{y} =-K\, (1-\delta)\, y+Ky(1+x)\,,
\end{equation}
which coincide with the (rescaled) rate equations~(\ref{SIdot}). Disease fade-out demands
a fluctuational orbit, for which the momenta $p_x$ and $p_y$ are nonzero. Before dealing with such  orbits,
consider the fixed points of the zero-energy Hamiltonian. There are exactly three such points,
all of them 4d saddles \cite{KM}. Two of them, $B=[0,0,0,\ln(1-\delta)]$
and $C=[0,0,0,0]$, describe infection-free steady states. Point $C$ is fluctuationless:
it corresponds to the saddle point of the rate equations. Point $B$ is fluctuational, as
its $p_y \neq 0$.  Finally, the fluctuationless fixed
point $A=[-\delta,(\delta/K)(1-\delta)^{-1},0,0]$ corresponds to
the endemic fixed point of the rate equations.

\begin{figure}[ht]
\includegraphics[width=2.3in,clip=]{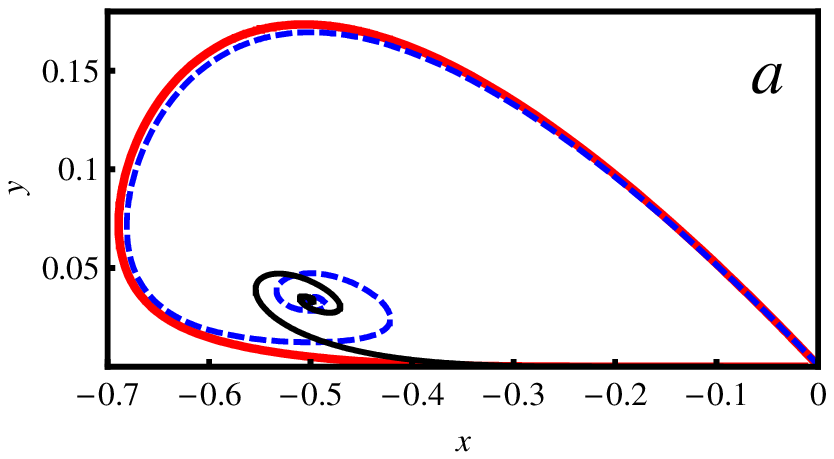}
\includegraphics[width=2.3in,clip=]{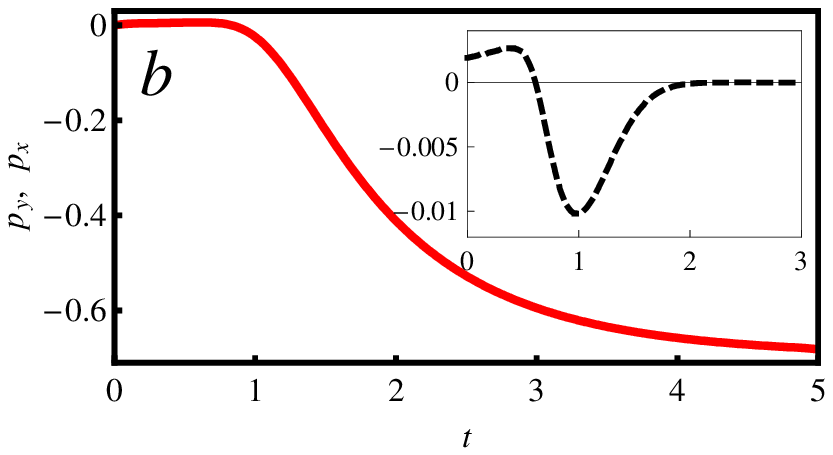}
\caption{(color online). \textit{a}: An epidemic outbreak on the $xy$-plane as predicted
by the rate equations~(\ref{foi_970}) (dashed line)
and the epidemic fade-out instanton  (thick solid line). Also shown
is the \textit{endemic} fade-out instanton \cite{KM} (thin solid line).  \textit{b}:
$p_y$ (inset: $p_x$) vs. $t$  for the epidemic fade-out instanton.
The rescaled parameters  are $K=30$ and $\delta=1/2$.}
\label{far}
\end{figure}

Let one or few infected  be introduced into an infection-free population.
In the leading WKB order this initial condition can correspond to different phase-space points
whose projections on the $xy$-plane
are very close to the fluctuationless fixed point $C$. Each of these phase-space points generates
an orbit which exits the fixed point $C$ along the manifold
spanned by it two
unstable eigenvectors. For epidemic fade-out to occur, such an orbit
must reach the extinction hyperplane $y=0$ before crossing, say, the hyperplane
$y=-(x/K) (1-\delta)^{-1}, \,-\delta<x<0$ (which is a 4d extension of the  artificial
absorbing boundary mentioned above). One can prove that, among all such orbits,  the one
providing the global minimum to the action $S(x,0)$ (and therefore the global maximum to the
fade-out probability  ${\cal P}_n$) ends in the fluctuational fixed point $B$. As a result,
$\max {\cal P}_n = {\cal P}_N$. Therefore, at $N\gg 1$, the epidemic fade-out problem
reduces to that of finding a heteroclinic orbit going from $C$ to $B$.
We found numerically
that such a heteroclinic orbit CB exists if and only
if $K>K_c=(1/4\delta) (1-\delta)^{-2}$:
when the endemic fixed point,
predicted by the rate equations, is a focus. As $K$ exceeds
$K_c$, the heteroclinic orbit emerges via a global bifurcation. In fact,
one finds \textit{multiple} heteroclinic orbits at $K>K_c$. They can be classified by whether their
$xy$-projections exhibit a single loop, two loops, three loops, \textit{etc}.
A single-loop
orbit, see Fig. \ref{far},  corresponds to disease fade-out immediately
after the first outbreak.   A two-loop orbit  corresponds to a fade-out immediately
after the second outbreak, \textit{etc}. The connection between the \textit{rapid} epidemic fade-out in a
stochastic population and a zero-energy instanton-like orbit of an effective Hamiltonian
is a central result of our work. We stress that previous studies, which found instanton-like
orbits in the context of population extinction, dealt with extinction from \textit{long-lived metastable} states.

\begin{figure}[ht]
\includegraphics[width=1.9in,clip=]{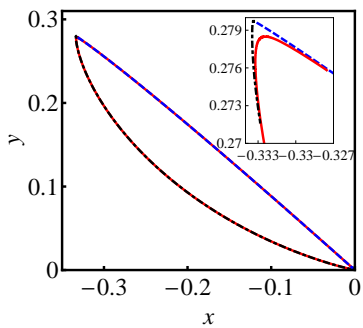}
\caption{(color online). Same as in Fig.~\ref{far} but for
$K=1.7875$ and $\delta=1/3$, so $K_c=1.6875$.
The \textit{endemic} fade-out instanton \cite{KM} is shown by the dash-dotted line.
Inset: a blowup near
the endemic fixed point A.}
\label{close}
\end{figure}

How does the epidemic fade-out instanton look like at different parameters?
For $K \delta \gg 1$ the fraction of infected versus time, $y(t)$, first
rapidly grows and becomes large and then falls down to a
small value [see Fig. \ref{y(t)}, solid line],
closely following
the prediction from the rate equations.
Then $y(t)$ strongly
deviates from the deterministic path and rapidly goes to zero. The $x$-, $p_x$- and $p_y$-dynamics
for the same values of $K$ and $\delta$ are
depicted in Fig.~\ref{far}. On can see that a rapid deviation from the deterministic path
occurs around $x=-\delta$. Notably,
$|p_x|$ remains much smaller than unity \textit{everywhere}. $|p_y|$, however, is steadily growing and, at
non-small $\delta$, becomes
${\cal O}(1)$ as the instanton
approaches the fluctuational fixed point $B$. As a result, the vKSSE is invalid
for most of the small-$y$ region where disease extinction occurs.

Near the bifurcation, $0<K-K_c \ll K_c$, our numerics reveals an
intimate relation between the epidemic
fade-out instanton and two other zero-energy
heteroclinic orbits. The first is the \textit{deterministic} orbit which lies
in the $xy$-plane and goes from C to A. The second is
the \textit{endemic} fade-out instanton: a heteroclinic orbit which goes from A to B and
describes stochastic \textit{endemic} fade-out \cite{KM}.
The $xy$-projection of the epidemic fade-out
instanton initially
closely follows
the deterministic orbit CA, see Fig. \ref{close}.
The momenta $p_x$ and $p_y$ are very small here.  They slowly
build up and become important only when the $xy$-projection of the instanton reaches a close vicinity of
the endemic
point A. Here the projection of the epidemic fade-out
instanton departs from the deterministic orbit (see the inset of Fig. \ref{close})
and rapidly approaches
the projection of the \textit{endemic} fade-out instanton.

\begin{figure}[ht]
\includegraphics[width=2.3in,clip=]{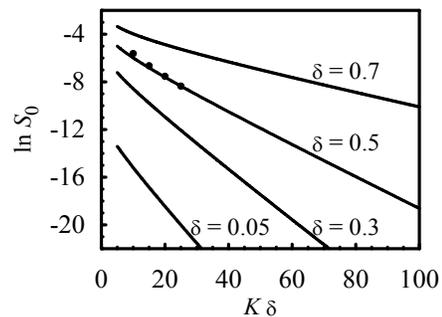}
\caption{Natural logarithm of the action ${\cal S}_0$ along the instanton
versus $K\delta$ at different $\delta$ as predicted by our asymptotic theory at $K \delta \gg 1$
(lines), and found by a numerical integration of the full Hamilton's equations (circles) for
$\delta=0.5$).}
\label{compare}
\end{figure}

To evaluate ${\cal P}\sim {\cal P}_N$ in the leading WKB order,
we need to calculate the accumulated action ${\cal S}_0$
along the instanton. In the rest of this communication we will focus on the important regime of $K\delta \gg 1$,
when the fade-out probability can indeed be significant. It turns out that the presence of the
small parameter
$({\cal K \delta})^{-1}$ enables one to find the instanton, and
calculate ${\cal S}_0$,
analytically.  An immediate simplification comes from the fact that the
fluctuations of the number of susceptibles are negligible everywhere, so
we can Taylor-expand the Hamiltonian (\ref{foi_40}) in $p_x \ll 1$ and truncate
the expansion at first order. Another simplification employs the strong inequality $y\ll \delta$ which
holds in the whole
region where the fade-out instanton significantly deviates from the deterministic orbit.
A complete calculation
of the instanton  will be reported elsewhere. Here we will analytically calculate ${\cal S}_0$.
As can be verified \textit{a posteriori},
the main contribution to ${\cal S}_0$ comes from a narrow region $|x+\delta| \ll \delta$,
where the instanton rapidly
departs from the deterministic orbit. Furthermore, $|p_y| \ll 1$ in this narrow region, so one can Taylor-expand Eq.~(\ref{foi_40}) in $p_y$ and truncate the
expansion at $p_y^2$. Neglecting small terms, we can reduce the
Hamiltonian (\ref{foi_40}) to
\begin{equation}\label{simpleH}
    H(x,y,p_x,p_y)\simeq \delta\, p_x+Kyp_y\left[x+\delta +(1-\delta)p_y\right]\,.
\end{equation}
The reduced problem is integrable. There is no need
in the full solution, however, if one only needs to
evaluate ${\cal S}_0$.
The Hamilton's equation for $\dot{x}$ yields $x(t)=\delta(t-1)$, where the arbitrary constant is
fixed by choosing $x(t=0)=-\delta$.   The Hamilton's equation for $\dot{p}_y$ reads
\begin{equation}
\dot{p}_y=-Kp_y\left[x+\delta+(1-\delta)p_y\right]\,.
\label{py_10}
\end{equation}
Plugging here $x=\delta(t-1)$, we obtain an exactly soluble equation for $p_y(t)$.
The boundedness of $p_y(t)$
fixes the integration constant, and we obtain
\begin{equation}
p_y=\frac{1}{K(1-\delta)}\frac{d}{dt}\, \ln \int_t^\infty e^{-\frac{K\delta}{2}u^2}\, du\,.
\label{py_20}
\end{equation}
Now let us calculate $\dot{S}$ along the instanton:
$\dot{S} = p_x \dot{x}+ p_y \dot{y} = H+ K(1-\delta) y p_y^2 \equiv {\cal F}$,
where we have used $H=0$ and denoted ${\cal F}\equiv K(1-\delta) y p_y^2$.
Using the Hamilton's equations, we observe that
${\cal F}(t)$ obeys the equation $\dot{\cal F}= - K (x+\delta) {\cal F} = - K\delta \,t {\cal F}$.
Integration yields
\begin{equation}\label{calF}
    {\cal F}(t)\equiv K(1-\delta) y(t) p_y^2(t) = C \exp (-K \delta t^2/2)\,,
\end{equation}
where $C=const$. Therefore, $\dot{S} = C \exp (-K \delta t^2/2)$,
and
\begin{equation}\label{S1}
    {\cal S}_0 =
\int_{-\infty}^{\infty} \dot{S}\,dt = C\sqrt{\frac{2\pi}{K\delta}}\,.
\end{equation}
What is left is to find $C$. Importantly,
the \textit{deterministic} solution still holds in the region  of
$-x-\delta\ll \delta$ (or $-t \ll 1$). For $K \delta \gg 1$
the deterministic solution was found by van Herwaarden \cite{vH2}.
In the region of $-x-\delta\ll \delta$, Eqs.~(3.25 a-d) of van Herwaarden
can be simplified and rewritten,  in our notation, as
\begin{eqnarray}
  y(t) &=& y_m\, \exp (K \delta t^2/2)\,, \label{MF_50}\\
   y_m &=& y_m(K,\delta)=\frac{\left(\delta+x_m\right)x_m}{1+x_m}\,
  \left(\frac{-x_m}{\delta}\right)^{K\delta}\nonumber \\
  & \times & \exp
\left[K(x_m+\delta)-(1+x_m^{-1})\, Q_1(x_m)\right]\,,
\label{MF_20}
\end{eqnarray}
where $x_m=x_m(\delta)$ is the negative root of the equation $x_m=(1-\delta)\ln (1+x_m)$,
and $Q_1(x_m)$ is given by
\begin{eqnarray}
  Q_1(x_m) &=& \int\limits_0^{x_m}\biggl[\frac{s(s+\delta)}{(1+s)^2\left[s-(1-\delta)\ln (1+s)\right]} \nonumber\\
  &-& \frac{x_m}{(1+x_m)\, (s-x_m)}\biggr]\, ds\,.
\label{MF_40}
\end{eqnarray}
(For $\delta\to 0$ one obtains $x_m(\delta)\simeq -2\delta$ and $Q_1\simeq -4\delta$.)
In the region of $(K \delta)^{-1/2} \ll -x-\delta \ll \delta $ Eq.~(\ref{py_20}) becomes
\begin{equation}
p_y(t) =-(1-\delta)^{-1} [\delta/(2\pi K)]^{1/2}\,\exp(-K\delta t^2/2)\,.
\label{py_30}
\end{equation}
Using Eqs.~(\ref{calF}), (\ref{MF_50}) and (\ref{py_30}) in their joint validity region $(K \delta)^{-1/2} \ll -x-\delta \ll \delta $, we obtain $C=y_m \delta/[2 \pi (1-\delta)]$.
Putting everything together, we obtain the leading-order WKB result for the epidemic
fade-out probability: ${\cal P} \sim \exp(-N {\cal S}_0)$, where ${\cal S}_0$ is
given by Eq.~(\ref{S1}) 
and $y_m$ is given by Eqs.~(\ref{MF_20}) and (\ref{MF_40}).
Note that ${\cal S}_0$ is exponentially small in $K\delta\gg 1$, so the WKB result holds only for very large
$N$: $N {\cal S}_0 \gg 1$.  In Fig. \ref{compare}  our results for ${\cal S}_0$ are compared
with those obtained by a
numerical integration of the Hamilton's equations. For large $K\delta$
the agreement is very good. For $N {\cal S}_0 \lesssim 1$ the epidemic fade-out probability is large.

That truncation of $H$ at $p_y^2$ yields, at $K\delta \gg 1$,
an accurate leading-order result for ${\cal S}_0$ justifies the validity
of the vKSSE \textit{for calculating} ${\cal S}_0$. Indeed,
our leading-order result for ${\cal P}$ coincides with that obtained, by an entirely
different method, by van Herwaarden \cite{vH2} whose starting point was the
vKSSE. We reiterate, however, that the
vKSSE
is invalid in most of the small-$y$ region, while the full
WKB Hamiltonian (\ref{foi_40}) holds there. Only at $\delta \ll 1$, when
$|p_y| \ll 1$ on the \textit{whole} instanton, the vKSSE describes the instanton correctly.
In this case one obtains
${\cal S}_0=(2 \delta^5/\pi e^4 K)^{1/2} (e/2)^{-K \delta}$.

In summary, we have shown that rapid epidemic fade-out in
stochastic populations is
amenable to an accurate analysis via a WKB theory. We calculated
the fade-out probability and
established an unexpected connection between the rapid epidemic fade-out and an instanton-like
orbit of an underlying Hamiltonian.
The fade-out instanton should be observable in stochastic simulations of, and
actual data on, epidemics in small communities.

This work was supported by the Israel Science Foundation (Grant No. 408/08).

\end{document}